\documentclass[prl,showpacs,twocolumn,floatfix]{revtex4} 
\usepackage{epsfig} 
\usepackage{mathrsfs} 
\usepackage{amssymb} 
\usepackage{color} 
\usepackage{graphics}
\usepackage{psfrag}
\usepackage{graphicx} 
\newcommand{\ecen}{\end{center}} \newcommand{\btab}{\begin{tabular}} 
\newcommand{\etab}{\end{tabular}} \newcommand{\bdes}{\begin{description}} 
\newcommand{\edes}{\end{description}}  
 \newcommand{\beq}{\begin{equation}} 
\newcommand{\eeq}{\end{equation}} \newcommand{\bea}{\begin{eqnarray}} 
\newcommand{\eea}{\end{eqnarray}}  
  
\newcommand{\bary}{\begin{array}} \newcommand{\eary}{\end{array}} 
\newcommand{\benum}{\begin{enumerate}} 
\newcommand{\eenum}{\end{enumerate}} \newcommand{\bitem}{\begin{itemize}} 
\newcommand{\eitem}{\end{itemize}} 

\begin{document}


\title{Physics of interface: Mott insulator barrier sandwiched between two
metallic planes}


\author{Sanjay Gupta} \email[]{sanjay1@bose.res.in}
\affiliation{ S. N. Bose National Centre for Basic Sciences, Kolkata, India}
\author{Tribikram Gupta } \email[]{trbkrm01@yahoo.co.in}
\affiliation{Theoretical Physics Division, 
Indian Association for the Cultivation of Sciences, Kolkata, India}


\date{\today}

\begin{abstract}
We consider a heterostructure of a metal and a barrier with onsite correlation
at half filling using unrestricted Hartree Fock. We find that above a certain 
value of correlation strength in the barrier planes, the system is a Mott 
insulator, while below this value the system still behaves like a gapless
insulator. The energy spectrum is found to be very novel with the 
presence of multiple gaps. Thus the system remains non metallic for any 
finite value of correlation. 

\end{abstract}


\maketitle

\section{Introduction} Recently there has been a spurt in the study of 
multilayered heterostructures, fabricated out of strongly correlated materials.
A wide range of systems have been studied experimentally and theoretically\cite{Ohtomo, 
Thiel, Lee, Kancharla, Okamoto, Rosch, Krish}.
The interface of a band insulator and a Mott insulators can show metallic behavior
\cite{Ohtomo} or can even become superconducting \cite{Thiel}. 
As shown by Thiel et al., such interfaces can be manipulated by gate
voltages thereby opening the prospect for interesting novel devices. 

In this letter, we investigate the interface of a metal and a system with finite 
onsite correlation. For onsite correlation greater than a certain value the system 
opens a gap, which mimics the Mott insulator. What happens when we sandwich this 
system between two metallic planes? We find that the overall system behaves as an 
insulator not only in the region where there is a gap but also in the region where 
the system is gapless.  

The introduction of metallic planes induces multiple gaps in the energy spectrum for 
high values of onsite correlation $U$. This can be understood to be due to the 
emergence of vastly different types of sites both in plane and perpendicular to it.
The gapless region behaves as a disordered insulator, with the disorder coming from 
the induced inhomogenity in the underlying Coulomb landscape. 
There is a charge reconstruction across the planes for such a system as we vary $U$.
In the region of parameters where a gap opens up, this reconstruction is more vigorous.

The interface that we have studied has been 
studied by others\cite{Rosch, Krish} using inhomogeneous dynamical 
mean field theory\cite{Georges, 
Metzner,Potthoff,freericks_idmft}. 
However, in the IDMFT approach used above all sites in each of the planes are treated 
as identical(spatial variation is taken only along the z direction), 
and thus misses out on crucial in plane modulations, which is one of the crucial reasons 
for the opening up of multiple gaps in the spectrum. It also considers a paramagnetic solution 
for the magnetic sector, which is not the correct solution for an underlying cubic lattice 
that we study. 
 

We employ the method of unrestricted Hartree Fock to solve the problem. It
can capture the physics arising due to spatial variations both in and 
perpendicular to planes. 
It can handle the charge and magnetic sector on 
the same footing and treats correlations in a self consistent fashion. 
Recently we have shown the utility of this method in describing the metallic 
phase that arises in two dimensions as a result of competition between 
correlation and disorder\cite{Nandini,Gupta}.  

\subsection{Model and Method} 
The barrier planes are described by the single orbital Hubbard model
and the metallic planes by the non interacting tight-binding model.  
The Hamiltonian for the system is \begin{eqnarray} 
\mathcal{H}&=&-\sum_{ij\alpha\sigma}t_{ij}^\parallel 
c^\dagger_{i\alpha\sigma}c^{}_{j\alpha\sigma} - t \sum_{i\alpha\sigma} 
[c^\dagger_{i\alpha\sigma}c^{}_{i\alpha+1\sigma}+h.c.]\nonumber\\ 
&-&\mu\sum_{i\alpha\sigma}c^\dagger_{i\alpha\sigma}c^{}_{i\alpha\sigma} 
+\sum_{i\alpha}U_\alpha(n_{i\alpha\uparrow} 
-\frac{1}{2})(n_{i\alpha\downarrow}-\frac{1}{2}).\; \end{eqnarray} Here the 
label $\alpha$ indexes the planes, and the label $i$ indexes sites of the 
two-dimensional square lattice in each plane. The operator 
$c^\dagger_{i\alpha\sigma}$ ($c^{}_{i\alpha\sigma}$) creates (destroys) an 
electron of spin $\sigma$ at site $i$ on the plane $\alpha$. We set the 
in-plane hopping $t^{\parallel}$ to be nearest neighbor only, and equal to 
$t$, the hopping between planes, so that the lattice structure is that of a 
simple cubic lattice.  
We take $U_\alpha = U$ for the barrier planes, and zero for 
the metallic planes. The chemical potential $\mu$ is calculated by demanding 
that there be exactly N electrons in the problem. This is done by taking the 
 average of the N/2 th and the N/2 + 1 th energy level. We 
make no further assumption about the magnetic regime, and thus retain all 
spin indices in the formulas below. We label the barrier planes with $\alpha$ 
values from 1 to m. Thus $\alpha = 1,m$ correspond to the 
metallic layers and all the other $\alpha$ values correspond to the barrier planes.

\begin{figure}
\begin{center}
   \begin{tabular}{cc}
      \resizebox{39.5mm}{!}{\includegraphics{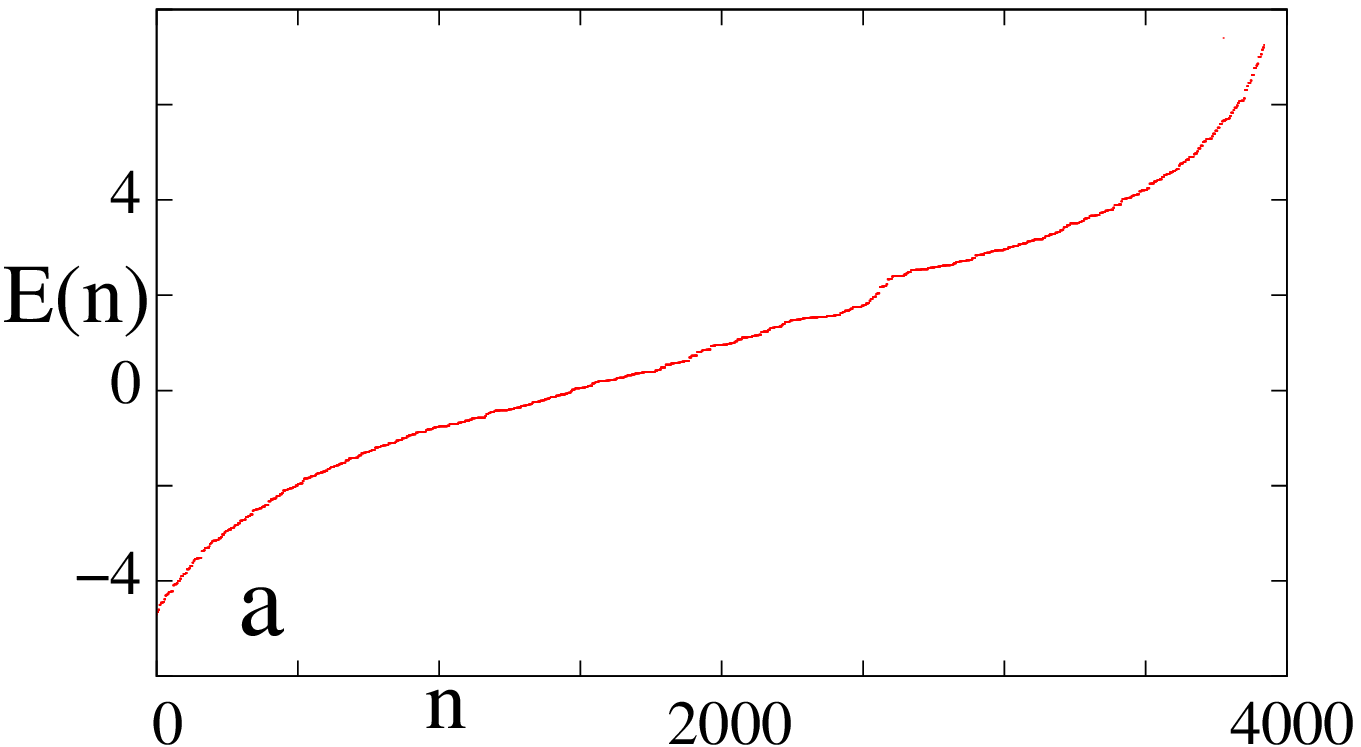}} &
      \resizebox{39.5mm}{!}{\includegraphics{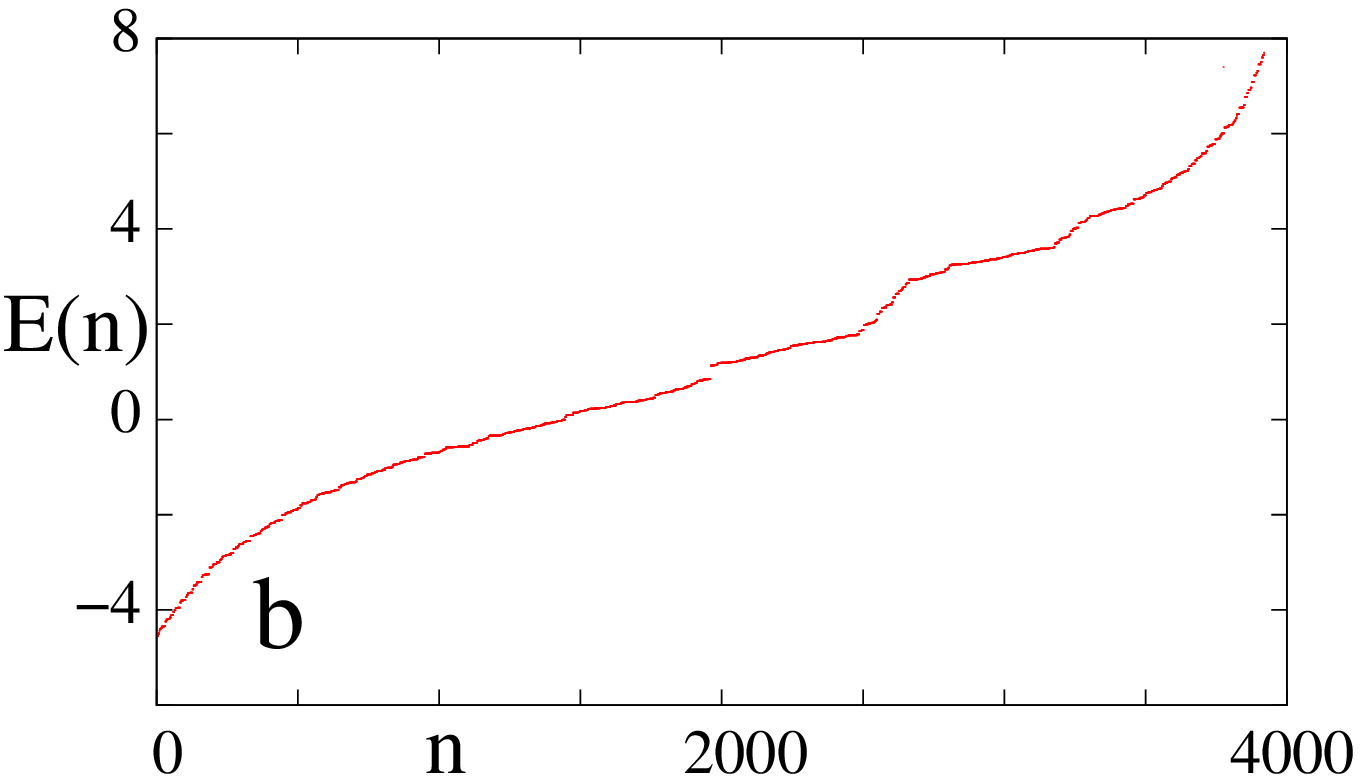}} \\ 
      \resizebox{39.5mm}{!}{\includegraphics{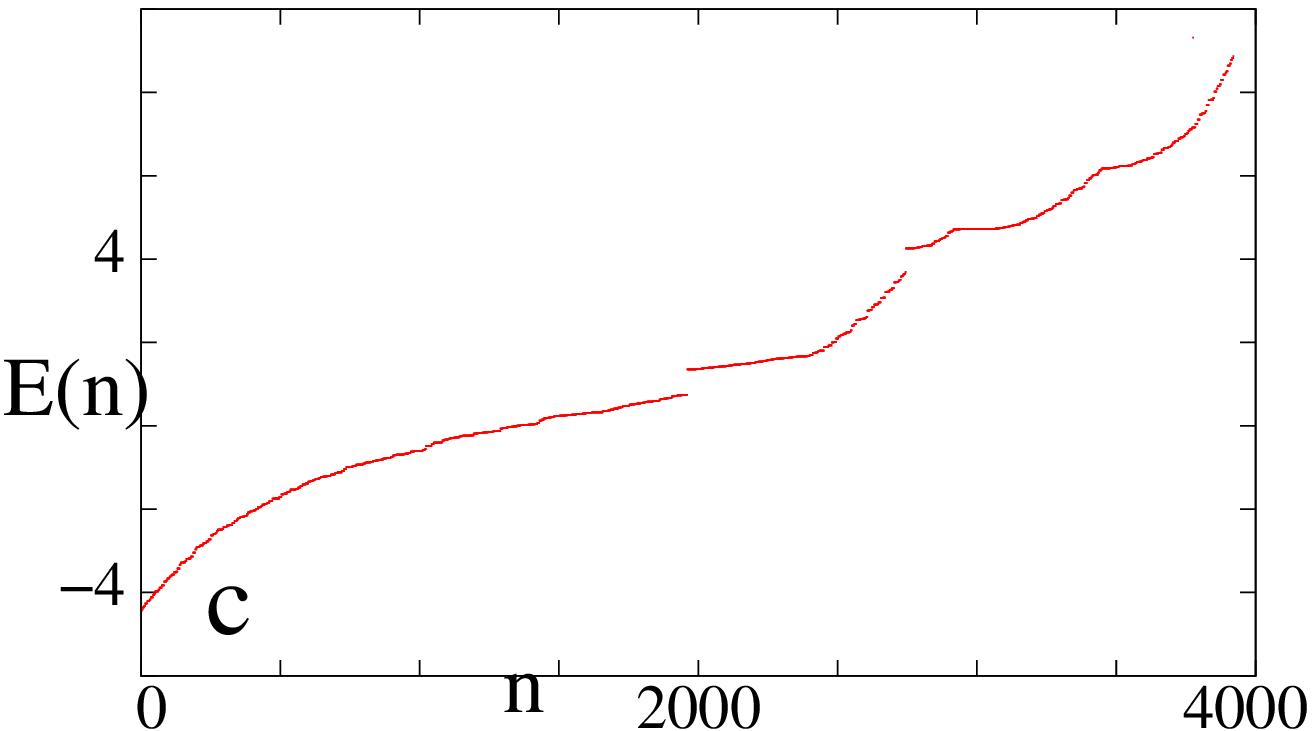}} &
      \resizebox{39.5mm}{!}{\includegraphics{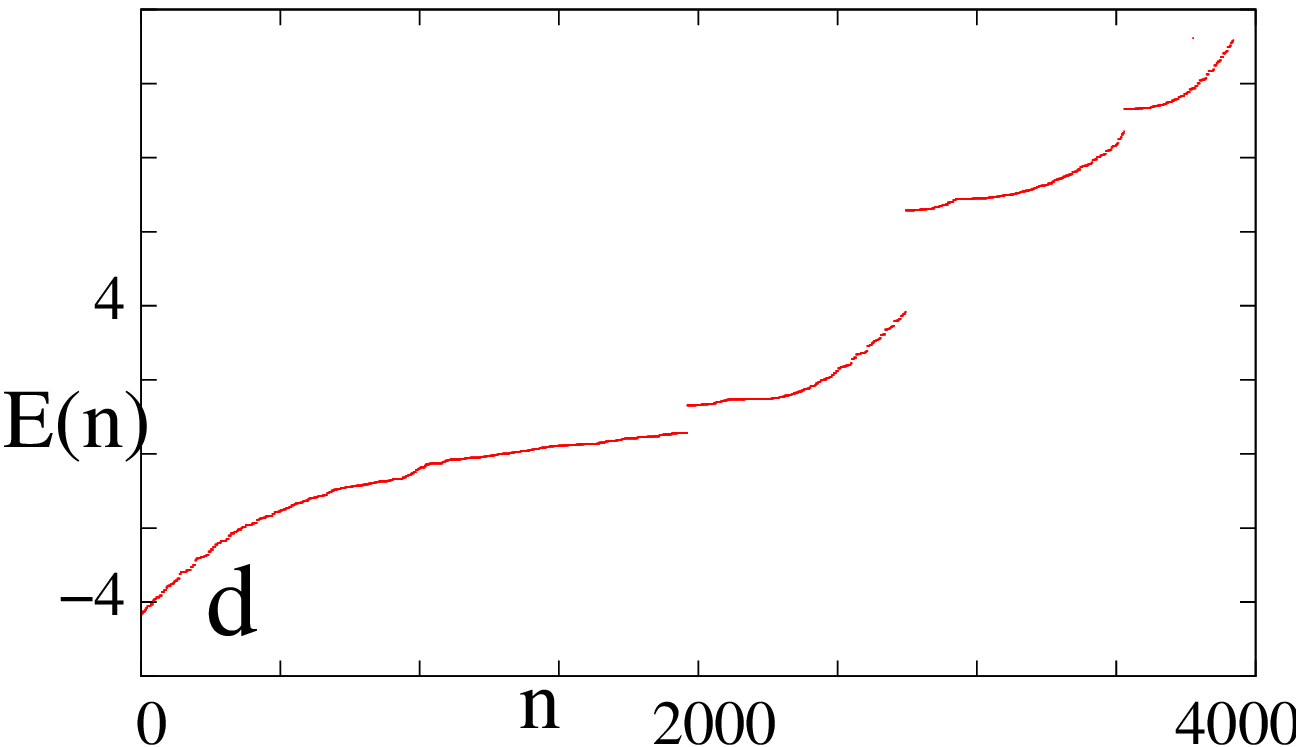}} \\ 
      \resizebox{39.5mm}{!}{\includegraphics{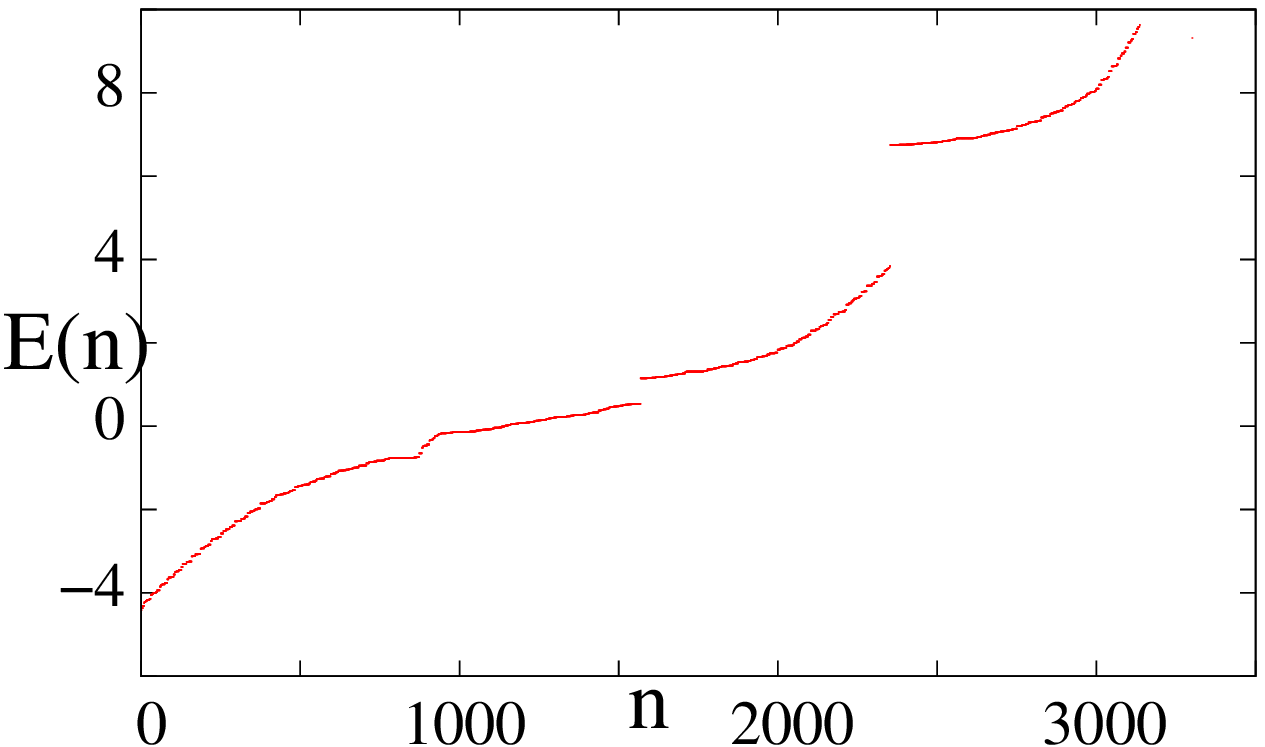}} &
      \resizebox{39.5mm}{!}{\includegraphics{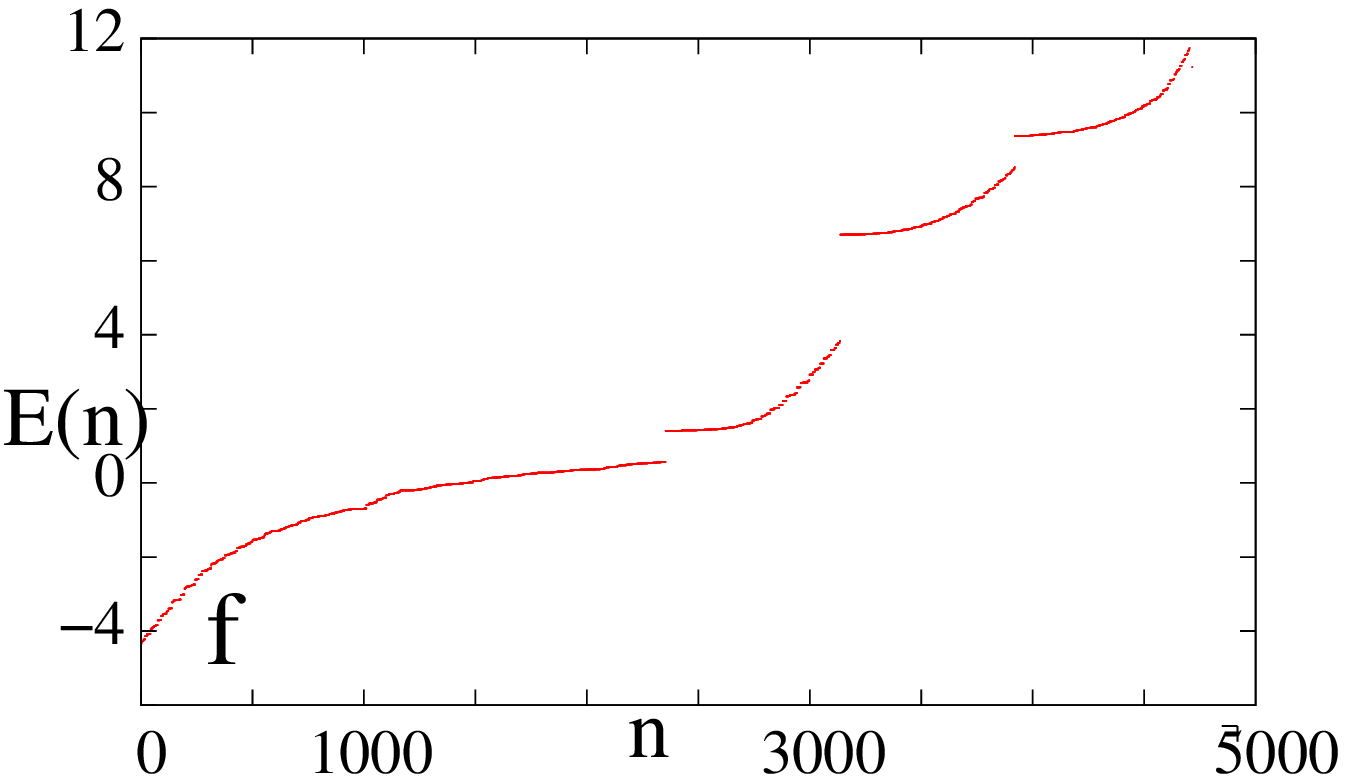}} \\ 
    \end{tabular}
\caption{Top left and right: $U = 4$ and $U =5$ for $28\times28\times5$ respectively.
Middle left and right: $U = 7$ and $U =10$ for $28\times28\times5$ respectively.
Bottom left and right: $U =10$ for $28\times28\times4$ and $28\times28\times6$ respectively.}
\label{fig;MultiGapatHalfFilling}
\end{center}
\end{figure}

\subsection{Calculated quantities} 
The energy spectrum of the system is very rich,  
with the presence of multiple gaps in it. We have 
also calculated the charge profile, spin profile and the dc 
conductivity at zero temperature. We calculate the effect 
of correlations and layer variation on the multiple gaps. 
The value of charge and spin at a particular site in the central square of 
each plane is taken and is plotted against the plane index to clearly 
bring out the inhomogeneous profile along the $z$ direction.
The charge at a particular site is simply calculated as 
$C$ = $n_{i,\uparrow}$ + $n_{i,\downarrow}$. The spin at a 
particular site is given by $S$ = $|(n_{i,\uparrow} - n_{i,\downarrow})|$.
The dc conductivity is calculated using the 
Kubo formula, which at any temperature is given by:
\begin{equation}
\label{eq.8}
	\sigma ( \omega)
	= {A \over N}
	\sum_{\alpha, \beta}(n_{\alpha}-n_{\beta})
	{ {\vert f_{\alpha \beta} \vert^2} \over {\epsilon_{\beta}
	- \epsilon_{\alpha}}}
	\delta(\omega - (\epsilon_{\beta} - \epsilon_{\alpha}))
\end{equation}
with $A = {\pi  e^2 }/{{\hbar a_0}}$, $a_0$ being the lattice spacing,
and $n_{\alpha}$ = Fermi function with energy $\epsilon_{\alpha}-\mu$.
The $f_{\alpha \beta}$ are matrix elements of the current operator
 $j_x = i t  \sum_{it, \sigma} (c^{\dagger}_{{i + x a_0},\sigma}
 c_{i, \sigma} - h.c)$, between exact single particle eigenstates
 $\vert \psi_{\alpha}\rangle$,
 $\vert \psi_{\beta}\rangle$, {\it  etc},  and
 $\epsilon_{\alpha}$, $\epsilon_{\beta}$
 are the corresponding eigenvalues. In this paper, conductivity/conductance
 is expressed in units of $A = {\pi  e^2 }/{{\hbar a_0}}$.

We calculate the `average' conductivity over small frequency intervals, 
$\Delta \omega$($\Delta \omega$ = $n \omega_{r}$, n = 1,2,3,4),
and then differentiate the integrated conductivity to get $\sigma(\omega)$
at $\omega = n \omega_{r}$, n = 1,2,3\cite{SanjeevEPL}. We repeat the 
same calculation for each temperature slice upto $kT = 0.1$ The 
temperature dependence of the conductivity tells us about the underlying 
nature of the overall system.  

For the quasi 2D geometry taken by us $N = m\times{L^2}$, where $m$ is the total 
number of layers and L is the total number of sites along 
both x and y direction, in the numerical work. 
We have performed finite size scaling of each of our results. 
We have shown  the result of finite size scaling on the dc 
conductivity for a particular parameter set($U = 5$) 
in this paper due to paucity of space. 
The effect of varying the number of Mott layers has been studied 
in great detail in this paper. 

\begin{figure}
\begin{center}
   \begin{tabular}{cc}
      \resizebox{39.5mm}{!}{\includegraphics{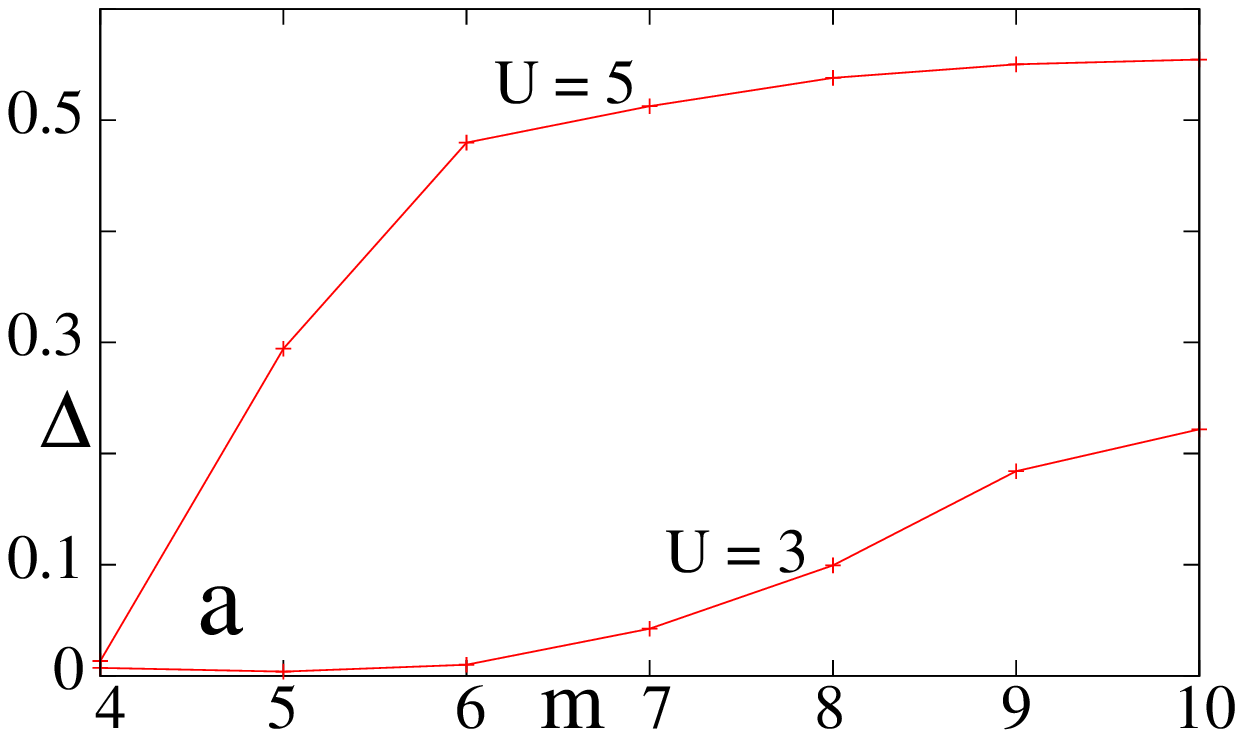}} &
      \resizebox{39.5mm}{!}{\includegraphics{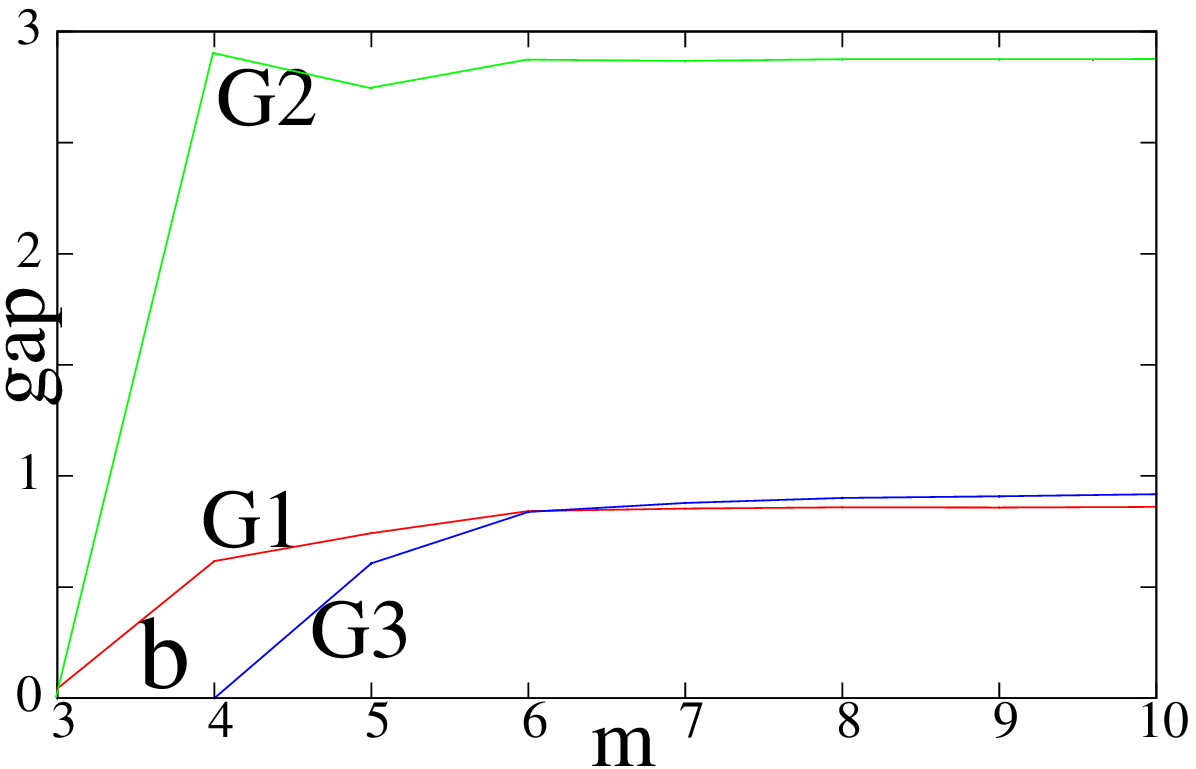}} \\ 
    \end{tabular}
\caption{left:Gap at half-filling with increasing Mott Layers for $U$ = 3 and 5
right:Effect of increasing number of layers on the multiple gaps.}
\label{fig;Gap_halffilling}
\end{center}
\end{figure}

\subsection{Our Work and Results}

We have calculated the dc conductivity $\sigma_{zz}(\omega_r)$, within the 
accuracy allowed by these finite size systems. The conductivity is 
calculated in units of the universal conductance $e^2/h$. 
Our method is able to capture the phenomenon of the gap opening as 
$U$ goes above a certain threshold. We find two types of insulators, 
one is gapped and the other is not. This second type of 
insulating behaviour which is shown for low $U$, $U \le U_{c}$,
($U_c$ is the value of $U$ above which the system opens up a gap), is 
a disordered insulator, where the disorder is introduced by the 
introduction of the metal interfaces at the bottom and top.

The effect of varying the insulating layer thickness on the charge 
order and spin order profile has been studied in this work.  
As the gap at half filling closes the 
spin order parameter collapses to near zero values. The charge order 
parameter on the other hand varies less dramatically across the gap 
closing transition. The charge profile builds up to 
its maximum value at the metallic planes and then dips to their lowest values in 
the first adjacent planes, from where there is massive charge depletion to the 
metallic planes, to gain Coulomb energy. As we move vertically deeper into the 
Mott planes, the total charge again picks up, reaching its bulk maxima at the 
central layer. There is overall symmetry around the central plane, as expected. 
As we increase the bulk thickness, we find that the charge 
peak at the central plane slowly approaches the value without the metallic 
planes, which is unity. This clearly indicates that the bulk becomes gradually 
insensitive to the metallic planes at the surfaces, as we increase the bulk layer 
thickness.

\begin{figure}
\begin{center}
   \begin{tabular}{cc}
      \resizebox{39.5mm}{!}{\includegraphics{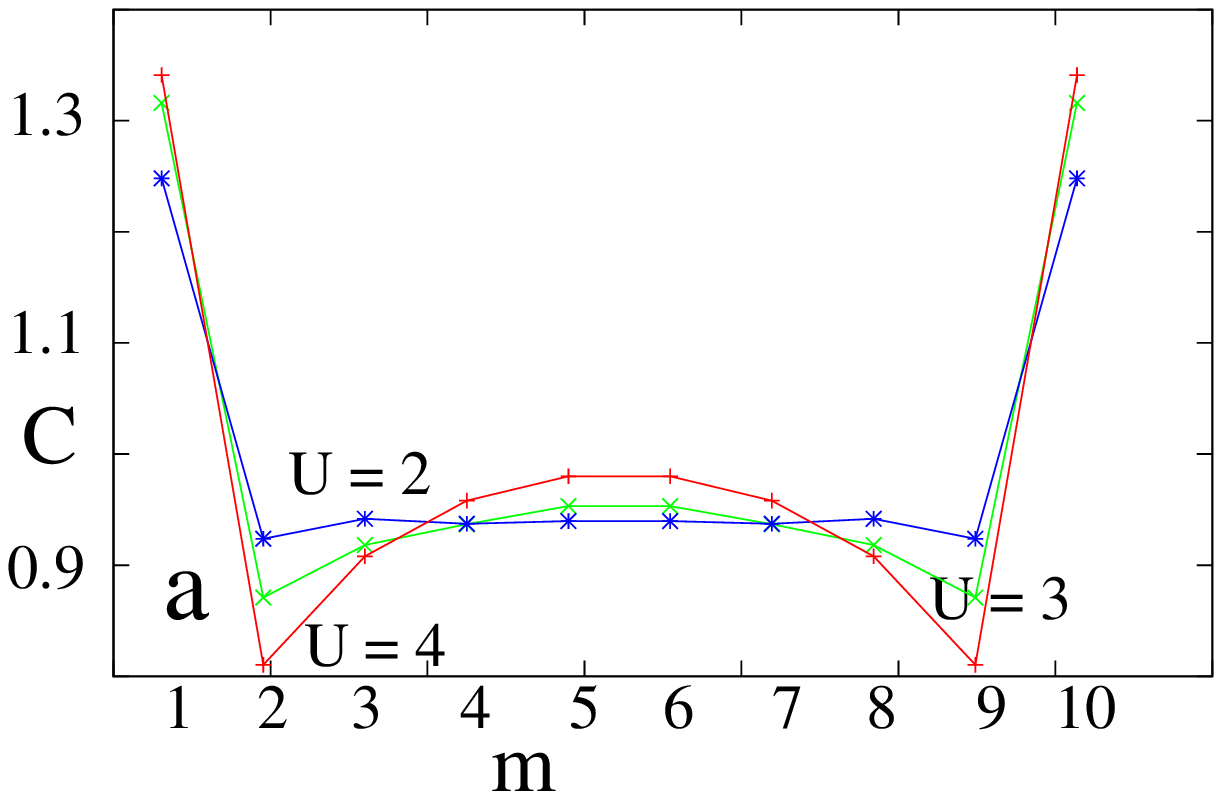}} &
      \resizebox{39.5mm}{!}{\includegraphics{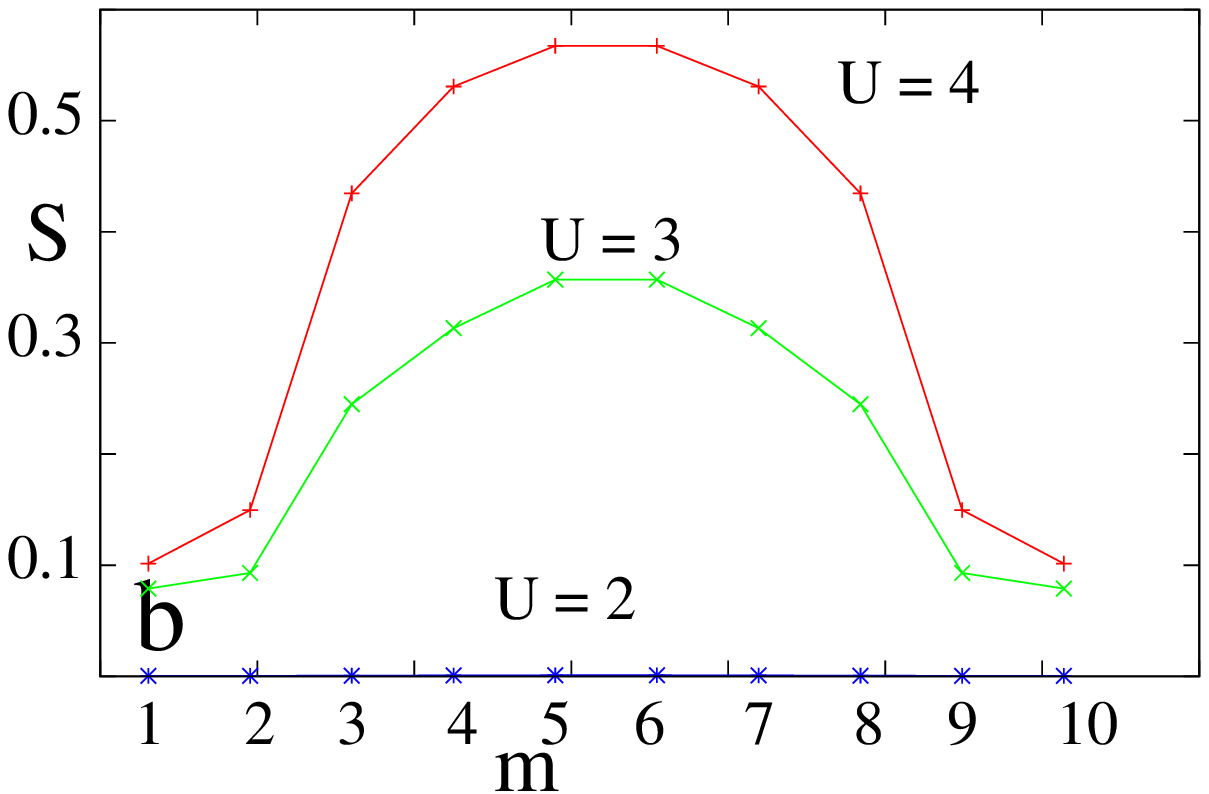}} \\
      \resizebox{39.5mm}{!}{\includegraphics{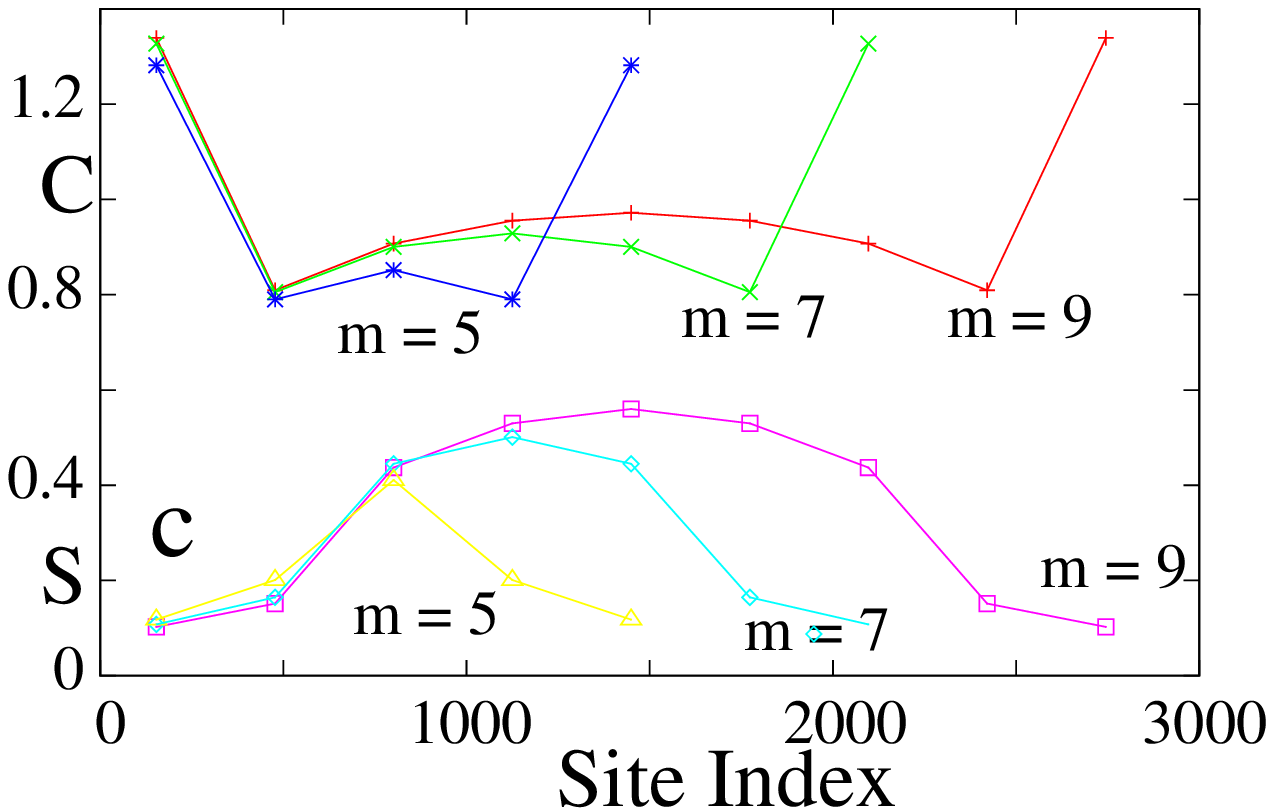}} &
      \resizebox{39.5mm}{!}{\includegraphics{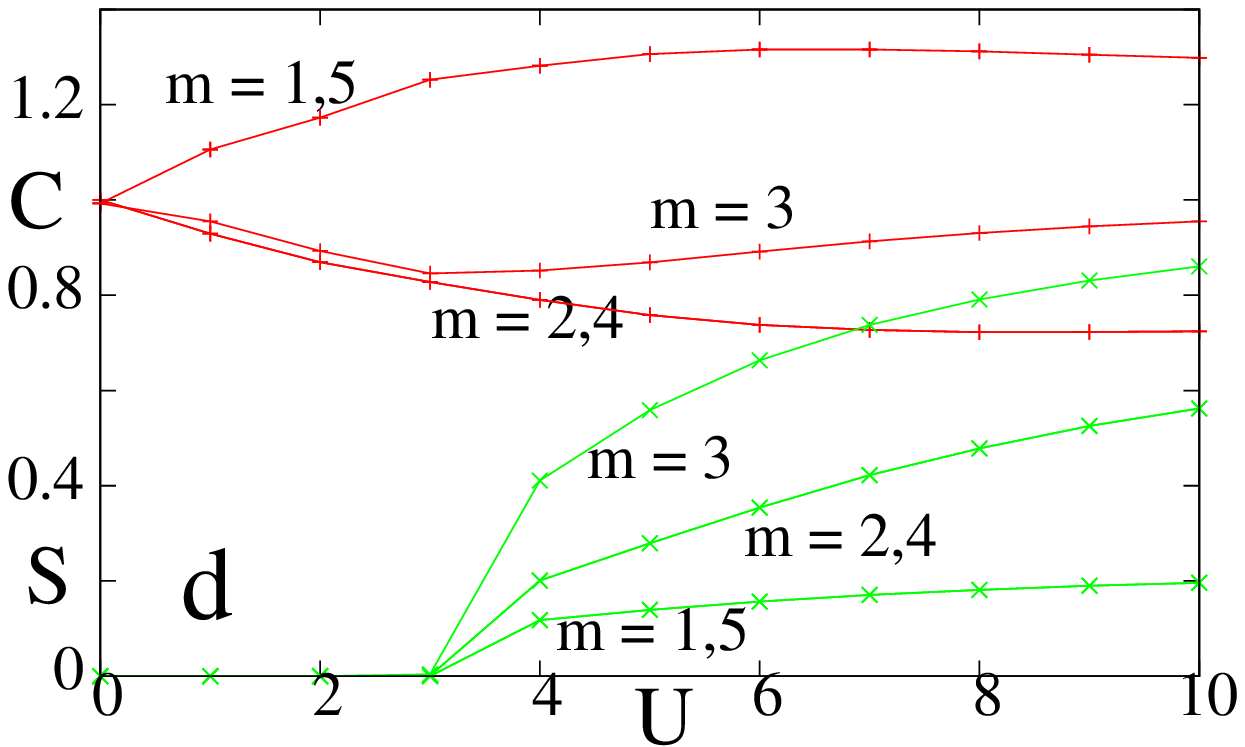}} \\
    \end{tabular}
\caption{Clockwise from top left:a Charge order variation vs layer index for a $18\times18\times10$ 
system, for $U$ = 2,3 and 4. 
b. Spin order parameter vs layer index for a $18\times18\times10$ system for $U$ = 2,3,4.
c. Spin and charge order vs central site index for $U = 4$ 
as the system is squeezed along the z direction.     
d. Spin and charge order vs $U$ for a $18\times18\times5$ system.}
\label{fig;Spin_ChargeOrder}
\end{center}
\end{figure}

\subsection{Analysis of our results} 
The sites having higher energy will have lower occupation and vice versa.
Fig 1a,b,c,d,e and f shows that different types of sites emerge in the problem 
that we study, as we systematically increase $U$. For different values of $U$, 
these gaps will open up successively. 
For $m = 5$, which means 3 intervening Mott layers, the first gap opens up 
roughly between $U = 4$ and $U = 5$, the second gap opens up close to $U = 7$ 
and the third gap opens up close to $U = 10$. The presence of 3 gaps clearly 
indicates the presence of broadly 4 different types of sites for $U = 10$. 

The 4 different types of sites correspond to the following. The first gap 
exactly at half filling corresponds to the situation when the $U$ has become 
large enough to just introduce some anti ferromagnetic ordering(however weak) 
even in the metallic planes. Thus the overall system develops a rather 
inhomogeneous Neel ordering, where the inhomogeneity is along the z direction. 
Another way of looking at it is to say that in the absence of metallic 
planes, there would be a big gap at half filling. The introduction of 
metallic planes reduces this gap to a somewhat lower value. As a result of this 
induction of anti ferromagnetic ordering(this will be discussed in more details
while explaining our spin/charge order results), two different types 
of sites emerge in the metallic plane for either spin.   
The reason for the generation of these two different type of sites in the metallic planes is 
as follows. 

To consider the emergence of two distinct types of sites due to the 
presence of interface, let us look at a particular spin(up/down). 
In the first type, an up electron sees comparatively high density of up spin and low density of down spin 
in the adjacent site in the Mott plane.  Such an up electron will experience  
lower Hubbard repulsion and greater Pauli blocking if it wants to hop to the adjacent site in the Mott plane. 
An up electron in the second type of site sees 
comparatively lower up electron density and higher down spin 
density in the adjacent site on the Mott plane.
This will lead to the second type of up electron 
experiencing higher Hubbard repulsion and 
lower Pauli blocking if it wants to gain kinetic energy     
by hopping to the adjacent Mott plane. 

As we increase $U$ to about 7, the energy band of the electrons in the Mott planes 
decouple completely from the energy band of the electrons in the metallic planes, thus
creating the second gap in the spectrum. The Mott layers now form the 
topmost band. As we crank up the $U$ further, the environment seen by the sites in the Mott 
planes gets split further. An electron located in the central plane/s has the strongest Neel 
order, while an electron located in the the Mott layers adjacent to the metallic planes 
have a relatively weaker Neel order. With increasing  $U$ the central layer becomes distinct 
from the two Mott layers which are adjacent to the metallic planes. 
This leads to the further splitting of the uppermost band at around $U = 10$.
In Fig. 1e for system size $28x28x4$, the uppermost gap is absent because of non
existence of the central layer.

Fig. 2a shows the plot of the gap at half filling as we increase the number of Mott 
layers for $U = 3$ and 5. It shows that for m = 4 and 5 there is no gap at half filling
till $U = 3$. The gap opens up when $U = 5$. 

Fig.2b shows how the multiple gaps in the energy spectrum  ($\Delta$) behaves on 
increasing the number of Mott layers for $U = 10$. This is because on increasing 
the width the number of sites at which the electron encounters $U$ will rise, so 
the barrier becomes stiffer. 

Fig 3a shows the plots of total charge at a particular point in the central square
of each layer vs the layer index for $m = 10$, for three different values of $U = 
2,3,4$. We can see clearly how as $U$ increases the charge profile in 
the Mott layers become more and more non uniform along the z direction, thus 
developing a more pronounced hump. The charge depletion from the 
Mott layers just adjacent to the metallic planes and the 
the charge accumulation in the metallic planes increases with increasing $U$.  
In Fig 3b we have shown the plot of $S$ vs layer index for $m = 10$ and $U$ = 2,3,4. 
While $S$ is almost zero and featureless for $U = 2$, it starts showing a prominent 
feature which gets sharper as $U$ is increased to 3 and then 4. 
Fig 3c shows the plots of $C$ and $S$ for $U = 4$ but for 
three different values of $m = 5, 7, 9$.  As we increase the 
number of Mott layers, for a fixed $U = 4$, $S$ becomes more pronounced,
while $C$ becomes more asymmetric in terms of increasing difference 
between the value in the central Mott plane and the Mott plane adjacent to the 
metal planes.

\begin{figure}
\begin{center}
   \begin{tabular}{cc}
      \resizebox{39.5mm}{!}{\includegraphics{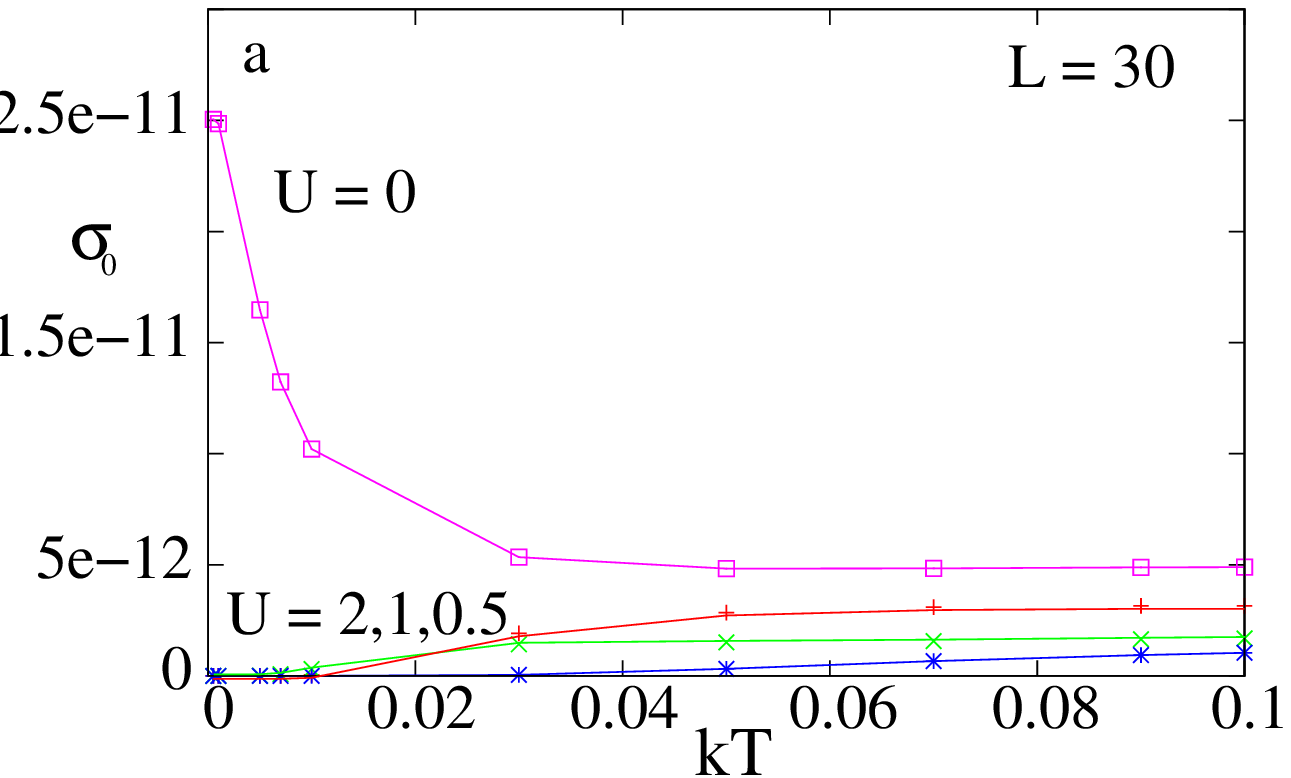}} &
      \resizebox{39.5mm}{!}{\includegraphics{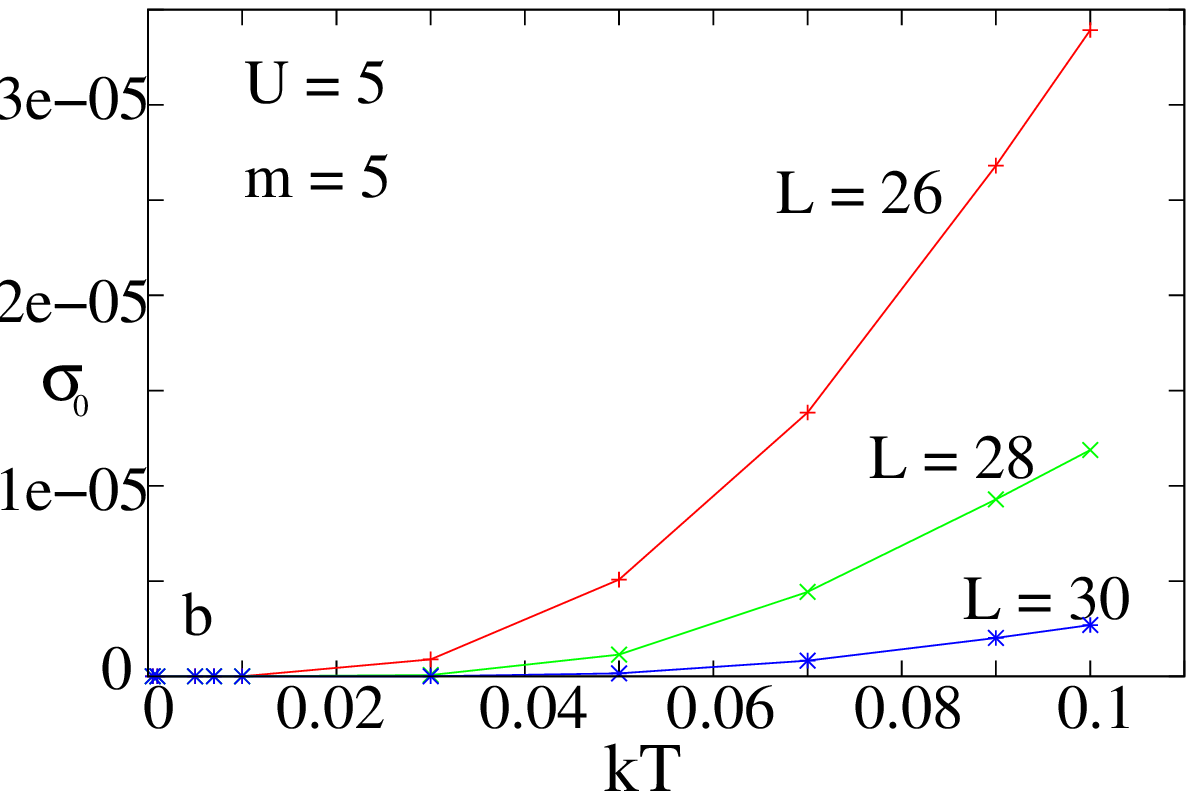}} \\ 
    \end{tabular} 
\caption{ Left: $\sigma_0$ vs $kT$ for $30\times30\times5$
system for $U$ = 0,0.5,1,2. The qualitative difference in the curves as the system goes from metal to 
insulator with increasing $U$ is captured very clearly. 
Right:Finite size scaling of $\sigma_0$ vs $kT$ for $m = 5$ and $U = 5$, for L=26,28 and 30.} 
\label{fig;CondvskT} 
\end{center} 
\end{figure}

In Fig 3d we show the plot of $C$ and $S$ vs 
$U$ for $18\times18\times5$ system. We have gone from $U$ = 0 to 
$U$ = 10 in integral steps. It clearly shows the emergence of 
three grossly different types of planes as we increase $U$. The 
three different types of planes are a) the two metallic planes, b) the two Mott 
planes adjacent to the metallic planes and c) the remaining Mott plane/s. 
We can see that as we increase $U$, more charge accumulates on the 
metallic planes and there is acute charge depletion from Mott planes adjacent
to the metallic planes. There is charge depletion from central Mott plane/s 
also initially, but as we increase $U$ further, the charge at the central 
Mott plane/planes increases slightly. This is accompanied by a large anti ferromagnetic 
ordering in the central planes/s. Thus the charges in the central Mott plane/s are 
unable to delocalize as it faces strong Coulomb repulsion on all sides and thus 
it piles up.  

In Fig. 4a we show the plots of dc conductivity against temperature $T$
for $U$ = 0,0.5,1 and 2 for a $30\times30\times5$ system,
which is the largest system size that we have considered. 
The $\omega_r$ has been chosen to be twice the 
finite size spacing between energy levels for each of the system size that 
we have considered. 
The dc conductivity curve for $U = 0$, shows metallic 
behaviour as expected with a sharply falling profile with increasing $T$.    
We note that the value of conductivity obtained for the metal is severely suppressed.
This is because, the highest contribution for the metal is at $\omega = 0$, 
which cannot be sampled in our finite size simulation. Thus we are able to 
sample the $\sigma(\omega)$ at some low but finite $\omega_r$ where the 
conductivity has already fallen very sharply.  
As we increase $U$ the dc conductivity for low 
$T$ gets severely suppressed by several orders of magnitude and the system 
is an insulator even for $U = 0.5$. 
This is due to the emergence of significant 
amount of inhomogeneity in the charge density landscape at $U = 0.5$. 
As we increase $U$ further, the amount of induced disorder increases 
further, suppressing conductivity even further. For $U = 5$, a small 
gap at half filling opens up, which seperates the occupied and unoccupied 
states.  Fig 4b shows the 
finite size scaling effect on the $\sigma(\omega_r)$ vs $T$ curves 
for $U = 5$, where we have taken $L$ = 26,28,30, all of which show 
insulating behaviour. 
We have convinced ourselves that the result of finite size scaling  
on the gap at half filling, converges to a fixed value for $L \ge 18$  
for $m = 5$. We find very high thermally activated conductivity
which increases by several orders of magnitude.

\section{Conclusion}
A heterostructure system arising due to the  
sandwich of a barrier with finite width and onsite correlations 
between two metallic planes is studied. There are multiple gaps 
in the spectrum, with increasing $U$. An inhomogeneous 
spin and charge profile develops in the system.
Dc conductivity calculations have been performed showing that the system 
is an insulator both in the gapped and gapless region.

\section{Acknowledgement}
Sanjay Gupta would like to thank the DST for providing the grant under the project "Electronic states and transport
 in mesoscopic/nanoscopic systems". 


\end{document}